\documentclass[11pt]{article}
\pdfoutput=1

\usepackage{euscript}
\usepackage{amsfonts}
\usepackage{amsbsy}
\usepackage{epsfig}
\usepackage{amsthm}
\usepackage{amscd}
\usepackage{amstext}
\usepackage{verbatim}
\usepackage{cancel}
\usepackage{capt-of}
\usepackage{empheq}

\usepackage[nosort]{cite}
\usepackage{bm}
\usepackage{authblk}
\usepackage{esint}
\usepackage[T1]{fontenc}
\usepackage{mathdots}
\usepackage[centertableaux]{ytableau}

\usepackage[utf8]{inputenc}
\usepackage{graphicx}
\usepackage{physics}
\usepackage{mathtools}
\usepackage{bbold}

\usepackage{setspace}
\usepackage{colortbl}
\usepackage{xcolor}

\usepackage{amsmath}
\makeatletter
\renewcommand*\env@matrix[1][\arraystretch]{%
  \edef\arraystretch{#1}%
  \hskip -\arraycolsep
  \let\@ifnextchar\new@ifnextchar
  \array{*\c@MaxMatrixCols c}}
\makeatother
\usepackage{amssymb}
\usepackage{mathrsfs}
\usepackage{booktabs}
\usepackage{bbm}
\usepackage{float}

\usepackage{pgfplots}
\pgfplotsset{compat=1.15}
\usepackage{tikz}
\usetikzlibrary{external}
\usetikzlibrary{arrows}

\usepackage{subcaption}
\usepackage{graphicx}
\usepackage{mathtools}
\usepackage[many]{tcolorbox}
\tcbset{shield externalize}
\usepackage{xcolor, colortbl}
\usepackage[a4paper]{geometry}
\usepackage[hidelinks,linktocpage]{hyperref}
\hypersetup{
    colorlinks=true,
    linkcolor=blue,
    citecolor=blue,
    }

\usepackage{mathrsfs}

\def\ben{\begin{equation}}
\def\een{\end{equation}}

\let\a=\alpha  \let\g=\gamma  
   
\let\l=\lambda

\let\w=\omega    \let\L=\Lambda

\let\pa=\partial
\def\be{\begin{equation}}
\def\ee{\end{equation}}
\def\beq{\begin{equation}}
\def\eeq{\end{equation}}
\def\ba{\begin{array}}
\def\ea{\end{array}}

\def\dalemb#1#2{{\vbox{\hrule height .#2pt
       \hbox{\vrule width.#2pt height#1pt \kern#1pt
               \vrule width.#2pt}
       \hrule height.#2pt}}}

\newcommand{\bea}{\begin{eqnarray}}
\newcommand{\eea}{\end{eqnarray}}
\def\ep{{\epsilon}}
\def\vep{{\varepsilon}}

\makeatletter
\newcommand*\bigcdot{\mathpalette\bigcdot@{.5}}
\newcommand*\bigcdot@[2]{\mathbin{\vcenter{\hbox{\scalebox{#2}{$\m@th#1\bullet$}}}}}
\makeatother

\def\R{{{\mathbb R}}}

\def\Z{{{\mathbb Z}}}

\def\ocal{{\mathcal{O}}}

\def\bk{{\bf k}}
\def\bx{{\bf x}}

\title{Lieb-Robinson causality and non-Fermi liquids}
\date{\vspace{-5ex}}
\author{Subham Dutta Chowdhury$^\sharp$, Sean A. Hartnoll$^\flat$ and Aditya Hebbar$^\sharp$}
\affil{
{\it $^\sharp$ The Abdus Salam ICTP,
Strada Costiera 11, 34151, Trieste, Italy} \\
{\it $^\flat$ DAMTP, University of Cambridge, Cambridge CB3 0WA, UK}
}

\begin{document}

\maketitle

\begin{abstract}

Quantum mechanical lattice models with local, bounded interactions obey Lieb-Robinson causality. We show that this implies a domain of analyticity of the retarded Green's function $G^R(\omega,{\bf k})$ of local lattice operators as a function of complex frequency $\omega$ and momentum ${\bf k}$, similar to the lightcone analyticity property of relativistic field theories. Low-energy effective descriptions of the dynamics must be consistent with this microscopic analyticity constraint. We consider two canonical low-energy fermionic Green's functions describing non-Fermi liquids, the marginal Fermi liquid and the `Hertz liquid'. The pole in these Green's functions must be outside of the Lieb-Robinson domain of analyticity for all complex momenta captured by the low-energy theory. We show that this constraint upper bounds the magnitude of the dimensionless non-Fermi liquid coupling in certain Hertz liquids. We furthermore obtain, from causality, an upper bound on the magnitude $|G^R(\omega,{\bf k})|$ within the analytic domain. We use this bound to constrain the quasiparticle residue of the non-Fermi liquids.

\end{abstract}

\tableofcontents

\newpage

\section{Introduction}

Lightcone causality is, of course, a foundational principle in relativistic physics. 
The fact that local operators commute at spacelike separation implies analyticity properties of the S-matrix in relativistic quantum field theory \cite{Goldberger:1955zza, Goldberger:1955zz, Bros:1965kbd, Sommer:1970mr, Martin:1965jj}. See \cite{Correia:2020xtr} for a summary of known results. Analyticity is a core ingredient in bootstrap approaches to quantum fields, as it connects high energy and low energy dynamics \cite{Pham:1985cr, Ananthanarayan:1994hf, Adams:2006sv, Pennington:1994kc, deRham:2022hpx, Kruczenski:2022lot}. Retarded Green's functions in relativistic theories enjoy similar analyticity properties through the K\"all\'en-Lehmann representation, and these have been used to bound dynamics in states where Lorentz invariance is broken by, for example, a temperature, charge density or magnetic field \cite{Creminelli:2022onn, Heller:2022ejw, Heller:2023jtd, Creminelli:2024lhd, Delacretaz:2025ifh, Hui:2025aja, Herzog:2025ddq}.

The Lieb-Robinson bound is a beautiful result that extends the notion of a lightcone to local lattice models with bounded interactions \cite{Lieb:1972wy}. The starting point for this bound is that a localised operator evolves in time through repeated commutation with the lattice Hamiltonian. If the Hamiltonian is also local, then each commutation can only cause the operator to spread in space by a finite amount. The spatial growth of the operator is thereby bounded by the amount of time that has passed. A rigorous treatment of this effect --- we will be using the formulation in \cite{Hastings:2010vzr, Chen_2023} --- shows that the time-evolved operator has exponentially small norm outside of a lightcone defined by the Lieb-Robinson velocity. The Lieb-Robinson velocity $v$ is, up to numerical factors, the lattice spacing multiplied by the norm of the local terms in the Hamiltonian --- see e.g.~\eqref{eq:v} below. It was argued in \cite{Hartman:2017hhp} that Lieb-Robinson causality could be used, in an analogous way to relativistic causality, to bound low-energy observables.

In this paper we will obtain sharp causality bounds on low-energy observables arising from lattice models. Our first result provides the tools that we need. In (\ref{eq:cons}) and in Fig.~\ref{fig:an} below we establish a domain of analyticity of retarded Green's functions, as a function of frequency $\omega$ and momentum ${\bf k}$, in lattice models that are subject to a Lieb-Robinson bound. The domain restricts the imaginary parts of the frequency and momentum --- see (\ref{eq:im}) below. At low frequencies and momenta the domain of analyticity is the same as that obtained for relativistic theories, with $v$ the effective speed of light. However, at large imaginary momenta, greater than the inverse lattice spacing, the `lightcone' becomes exponentially weaker. It is still present, however.

Non-Fermi liquids are ubiquitous in strongly correlated materials, but continue to present a major theoretical challenge because they involve strong interactions between the electrons \cite{RevModPhys.94.041002, pwp, RevModPhys.94.035004}. Non-Fermi liquids are often defined as exhibiting unconventional transport or thermodynamic properties \cite{Schofield01031999, RevModPhys.73.797}, but the most direct probe of the underlying electronic dynamics is angle-resolved photoemission spectroscopy (ARPES) \cite{RevModPhys.75.473, RevModPhys.93.025006}. ARPES measurements reveal the electronic retarded Green's function $G^R(\omega,\bk)$, defined precisely in (\ref{eq:gr}) below. In this paper we will show that the analyticity properties of $G^R(\omega,\bk)$, following from Lieb-Robinson causality, constrain the space of possible non-Fermi liquid theories that can emerge at low frequencies.

This paper will follow the `quantum lattice bootstrap' methodology introduced in \cite{Chowdhury:2025dlx}. The logic is to start with an effective low-energy description as given, and ask whether there are microscopic lattice constraints on the parameters in the low-energy description. In \cite{Chowdhury:2025dlx} we obtained constraints from microscopic bounds on the spectral weight $\text{Im} \, G^R(\omega,\bk)$. Here we will obtain constraints from microscopic causality.

The most basic requirement is that any pole in the electronic Green's function $G^R(\omega,\bk)$ must remain outside of the Lieb-Robinson domain of analyticity. In \S\ref{sec:fermi} we review some canonical non-Fermi liquid Green's functions, including the marginal Fermi liquid and a phenomenological generalisation of Hertz theory that we call a `Hertz liquid'. These models are assumed to pertain below some cutoff energy $\Lambda$. The parameters in the models are then the `quasiparticle residue' ${\mathcal A}$, a dimensionless coupling $\lambda$ and a renormalised Fermi velocity $v_F^\star$. Clearly one expects $v_F^\star < v$, although as we note in \S\ref{sec:bounds} there is room for exceptions. Our second main result, in (\ref{eq:final}), is an upper bound on the coupling $|\lambda|$ in cases where the fermionic excitation remains well-defined at the lowest energy scales but has a non-Fermi liquid decay rate (i.e.~$\omega < |\Sigma''(\omega)| < \omega^2$ as $\omega \to 0$). Our bound is consistent with ARPES data in overdoped cuprates, as fit in \cite{Reber2019, Smit2024}. In cases with stronger departures from Fermi liquid theory at low energies all velocities are driven to zero at the lowest energies and there is no bound on $\lambda$ from causality.

In \S\ref{sec:res} we obtain, using Lieb-Robinson causality, the bound (\ref{eq:bbb}) on the magnitude $|G^R(\omega,\bk)|$ of the Green's function within the domain of analyticity. This bound then implies our third main result, the upper bound (\ref{eq:Ab}) on the quasiparticle residue ${\mathcal A}$. The bound is, again, strongest for cases where the quasiparticle peak remains well-defined at the lowest energy scales. Roughly, if the pole in the Green's function can come close to the domain of Lieb-Robinson analyticity, then the magnitude of the Green's function within the analytic domain becomes too large unless the residue is small. In the discussion \S\ref{sec:disc} we extend our framework to the retarded Green's function for the current operator. Within the Drude model of transport, we show that our upper bound on $|G^R(\omega,0)|$ now leads to an upper bound on the dc conductivity.

\section{Analyticity from Lieb-Robinson causality}

Let the operator $\psi^\dagger_\a$ create a fermion at the site $\a$ of a lattice, which is at position $\bx_\a$. We have suppressed any fermion spin label. The retarded Green's function of this fermion is
\be\label{eq:gr}
G^R(\omega,\bk) \equiv -i \sum_\a \int_0^\infty dt \, e^{i \omega t - i \bk \cdot \bx_\a } \left\langle \left\{\psi^\dagger_\a(t), \psi^{\phantom{\dagger}}_0(0) \right\} \right\rangle_T  \,,
\qquad \left\langle \; \cdot \;  \right\rangle_T \equiv \tr \left( \cdot \;\; \frac{e^{- H/T}}{{\mathcal Z}_T} \right) \,,
\ee
where ${\mathcal Z}_T \equiv \tr e^{-H/T}$ is the partition function, with $T$ the temperature and $H$ the Hamiltonian. 
The curly bracket is the anticommutator, as usual, and $\bk$ can be taken in the Brillouin zone.

In this section we will derive a domain of analyticity of the retarded Green's function (\ref{eq:gr}), viewed as a function of complex $\omega$ and $\bk$. This will follow from controlling the Fourier transform using the Lieb-Robinson bound, which we now recall. Firstly, define
\be\label{eq:operator}
C(t,\bx_\a) \equiv \norm{\left\{\psi_\a^\dagger(t),\psi^{\phantom{\dagger}}_0(0) \right\}} \,.
\ee
This is an operator norm, which upper bounds the correlation function,
\be\label{eq:tt}
\left| \left\langle \left\{\psi^\dagger_\a(t), \psi^{\phantom{\dagger}}_0(0) \right\} \right\rangle_T \right| \leq C(t,\bx_\a) \,.
\ee
The Lieb-Robinson bound is an upper bound on the operator norm. For concreteness, let us restrict attention to a square lattice in $d$ spatial dimensions with lattice spacing $a$. We will also consider, for simplicity, a translationally invariant nearest-neighbour Hamiltonian given by an operator $H$ on each edge. These assumptions are not essential. More complicated lattices or longer range interactions will only change the numerical value of the Lieb-Robinson velocity $v$ in (\ref{eq:v}) below. Note also that for a fermion creation operator 
$\norm{\psi^\dagger_0(0)  \psi^{\phantom{\dagger}}_0(0)} = 1$.
This immediately implies that $C(t,\bx) \leq 2$, with the factor of 2 from the anticommutator. The Lieb-Robinson bound can be formulated as the stronger constraint, derived in Appendix \ref{app:lr}, that
\be\label{eq:C2}
C(t,\bx) <
\left\{
\begin{array}{cl}
K \left(\frac{e h t}{x/a}\right)^{x/a} & \text{for $x/a > eht$}  \\
2 & \text{for $x/a < e h t$}
\end{array}
\right. \,.
\ee
Here $x \equiv |\bx|$ and we defined $h \equiv 4(2d-1) ||H||$ and $K \equiv \frac{2d}{2d-1}$. We see in (\ref{eq:C2}) that $C(t,\bx)$ is forced to decay exponentially outside the `lightcone' $x = v t$, where the Lieb-Robinson velocity
\be\label{eq:v}
v \equiv e h a \,.
\ee

Now consider the Green's function (\ref{eq:gr}) with complex arguments. Denote the imaginary parts
\be\label{eq:im}
\omega_2 \equiv \text{Im}\, \omega \,, \qquad {\bf q} \equiv \text{Im}\, \bk \,.
\ee
The imaginary part ${\bf q}$ of the momentum is not restricted to the Brillouin zone. We want to determine the set $\Gamma$ of $(\omega_2, {\bf q})$ such that
the Fourier transform in (\ref{eq:gr}) converges.
Convergence depends on the competition between the non-oscillatory exponential terms due to the imaginary parts $(\omega_2, {\bf q})$ and the exponential decay in (\ref{eq:C2}).
To this end, consider the absolute value
\begin{align}
\left| G^R(\omega,\bk) \right| & \leq \sum_\a \int_0^\infty dt \, e^{- \omega_2 t + q x_\a } \left| \left\langle \left\{\psi^\dagger_\a(t), \psi^{\phantom{\dagger}}_0(0) \right\} \right\rangle_T \right| \\
& < \sum_\a \left[K \int_0^{\frac{x_\a}{eah}} dt \, e^{- \omega_2 t + q x_\a + \frac{x_\a}{a} \log \frac{e a h t}{x_\a}} + 2 \int_{\frac{x_\a}{eah}}^\infty dt \, e^{- \omega_2 t + q x_\a} \right] \,.\label{eq:grab}
\end{align}
Here we used (\ref{eq:tt}) and (\ref{eq:C2}) in the second line. We set $q = |{\bf q}|$ and $x_\a = |{\bf x_a}|$ and used ${\bf q} \cdot {\bf x}_\a \leq q x_\a$. The convergence of (\ref{eq:grab}) is determined by the behaviour of the integrand at large $t$ and $x_\a$. In this regime we may think of $x_\a \to x$, a continuous variable. Consider, then,  the Fourier transform along the line
\be
t = \g \frac{x}{a} \,,
\ee
and look at the behaviour at large $x/a$ and fixed $\gamma$. The integrands in (\ref{eq:grab}) at large $x/a$ are
\be\label{eq:exp}
\left\{
\begin{array}{ll}
K \, e^{ - \left[ \omega_2 \gamma -  a q - \log \left(e h \gamma \right)  \right] \frac{x}{a}} & \text{if} \quad e h \gamma \leq 1\\[5pt]
2 \, e^{ - \left[ \omega_2 \gamma -  a q  \right] \frac{x}{a}} & \text{if} \quad e h \gamma \geq 1
\end{array}\right. \,.
\ee
For the integral to converge we need the exponents in (\ref{eq:exp}) to be negative for all $\gamma$.
Depending on $\omega_2$, either the first or the second line (\ref{eq:exp}) will impose the stronger constraint. In the first line, the minimum of the quantity in square brackets in the exponent is reached at $\gamma = 1/\omega_2$. Requiring the exponent to be negative at this value gives the constraint $\omega_2 > h e^{aq}$. This will be the strongest constraint on $\omega_2$ and $q$ when the minimum is within the domain of applicability of the first line, $e h \gamma \leq 1$. That is, this is the strongest constraint for $\omega_2 \geq e h$. If instead $\omega_2 \leq e h$ then the strongest constraint occurs at the boundary $e h \gamma = 1$. A negative exponent at this value requires $\omega_2 > e h a q$. Putting the above results together we find that the integral converges for
\be\label{eq:cons}
\left\{
\begin{array}{ll}
\omega_2 > h e^{aq } & \text{for} \quad \omega_2 \geq e h\\[5pt]
\omega_2 > e h a q & \text{for} \quad \omega_2 \leq e h
\end{array}\right. \,.
\ee

\begin{figure}[h]
\centering
\includegraphics[width=0.5\textwidth]{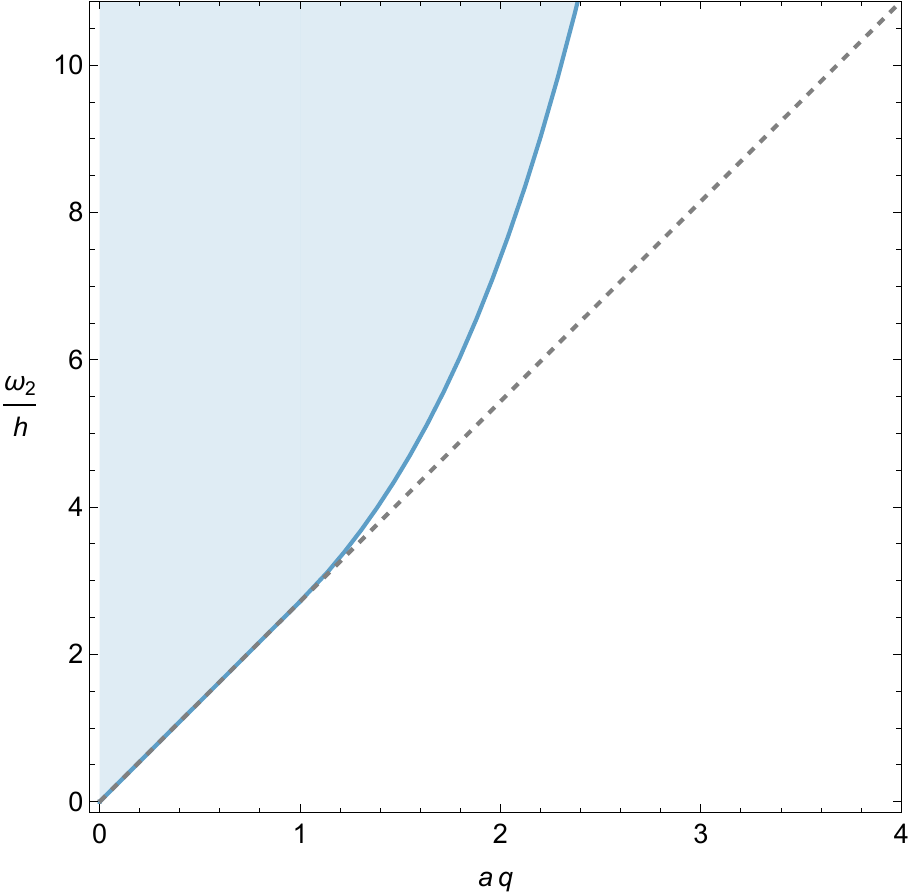}
\caption{The retarded Green's function must be analytic in the shaded region, given by (\ref{eq:cons}). Recall that $\omega_2$ and $q$ are the imaginary parts of the frequency and momentum as defined in (\ref{eq:im}). The dashed line shows the Lieb-Robinson lightcone with the velocity defined in (\ref{eq:v}).}
\label{fig:an}
\end{figure}

The retarded Green's function $G^R(\omega,\bk)$ is analytic in the domain of convergence $\R^{1+d} + i \Gamma$, with $\Gamma$ given by (\ref{eq:cons}). Analyticity holds because differentiating $G^R(\omega,\bk)$ with respect to $\omega$ or $\bk$ will bring down factors of $t$ or $x$ inside the integral that are subleading compared to the exponential suppression (\ref{eq:exp}) in the domain of convergence.\footnote{With nonvanishing imaginary parts inside the domain $\Gamma$, convergence and analyticity follow from (\ref{eq:grab}). For purely real frequencies and momenta, existence of the Fourier transform (\ref{eq:gr}) is the statement that the $(t,\bx_\a)$ Green's function is a tempered distribution. See \S2 of \cite{streater2000pct} for a more mathematically rigorous discussion.} This domain of analyticity is our first main result. The domain constrains the imaginary parts of the frequency and momentum, the real parts are unconstrained.  A plot of the region (\ref{eq:cons}) is shown in Fig.~\ref{fig:an}.
At low frequencies the bound defines a lightcone with the Lieb-Robinson velocity $v$ in (\ref{eq:v}). At frequencies above the microscopic energy $e h$ the bound (\ref{eq:cons}) becomes weaker and does not efficiently enforce analyticity at momenta well below the inverse lattice spacing.
We recall, however, that the imaginary part of the momentum is not constrained to the Brillouin zone. The bound in (\ref{eq:cons}) is therefore a refinement of the bound noted previously in Eq.~36 of \cite{Chowdhury:2025dlx}.

\section{Low energy dynamics close to a Fermi surface}
\label{sec:fermi}

Any effective description of the theory that pertains at low energies and long wavelengths must be compatible with the analyticity domain (\ref{eq:cons}). Let us now recall how to describe low energy fermionic excitations about a Fermi surface.

The full electronic retarded Green's function (\ref{eq:gr}) is expressed in terms of the self-energy $\Sigma$ as
\be\label{eq:full}
G^R(\omega,\bk) = \frac{1}{\omega - \epsilon_\bk -\Sigma(\omega,\bk)} \,.
\ee
We are working at zero temperature throughout. At real frequencies and momenta it is conventional to denote the real and imaginary parts of the self-energy as $\Sigma = \Sigma' + i \Sigma''$. The Fermi surface is then made up of points in momentum space obeying
\be\label{eq:kf}
\epsilon_\bk + \Sigma'(0,\bk) = 0 \,.
\ee
Note that $\Sigma''(0,\bk) = 0$ on these points \cite{PhysRev.119.1153, sachdev2023quantum}. This is the singular locus of the Green's function at zero energy. Low energy fermionic excitations therefore live close to the Fermi surface. To zoom in on these excitations we set
\be
\bk = \bk_F + \bk_\perp \,,
\ee
where $\bk_F$ is a point on the Fermi surface, obeying (\ref{eq:kf}), and $\bk_\perp$ is orthogonal to the Fermi surface. The effective theories we are about to describe arise by expanding the full Green's function (\ref{eq:full}) at small $\omega$ and $k_\perp \equiv |\bk_\perp|$. We will consider two classes of effective theories. These theories hold only at low frequencies and momenta, below cutoffs $\Lambda$ and $\bar \Lambda$,
\be\label{eq:bounds}
|\omega| \ll \Lambda \qquad \text{and} \qquad k_\perp \ll \bar \Lambda \,.
\ee

\subsection{Marginal Fermi liquid}

The marginal Fermi liquid (MFL) was originally proposed as a phenomenological model for the normal state of the cuprate high temperature superconductors \cite{PhysRevLett.63.1996}. For a recent discussion see \cite{RevModPhys.92.031001}. It captures widely observed physical properties of non-Fermi liquids, including a $T$-linear resistivity and logarithmic in temperature specific heat.

The low energy excitations are governed by the Green's function
\be\label{eq:gr2}
G_\text{MFL}^R(z,\bk) = \frac{{\mathcal A}}{z - v_F^\star k_\perp - \Sigma_\text{MFL}(z)} \,.
\ee
Here $v_F^\star$ is the renormalised Fermi velocity and we have used $z$ to emphasize that we are interested in complex frequencies. The self-energy takes the form
\be\label{eq:se1}
\Sigma_\text{MFL}(z) = \lambda \, z \log \left(\frac{- i z}{\Lambda}\right) + \ocal\left(\frac{z^2}{\Lambda} \right) \,,
\ee
where $\lambda \geq 0$ is the coupling and $\Lambda$ is the cutoff energy scale (\ref{eq:bounds}). In (\ref{eq:se1}) any additional linear-in-$z$ terms in the self-energy have been absorbed into a rescaling of the bare $z$ term, leading to the nontrivial weight ${\mathcal A} \neq 1$ in the numerator of (\ref{eq:gr2}). Higher order analytic corrections to the self-energy must be retained in  (\ref{eq:se1}) but, as indicated, are suppressed relative to the bare term by powers of $z/\Lambda$. We will drop these terms in the remainder. For real frequencies $\omega$:
\be
\Sigma_\text{MFL}'(\omega) = \lambda \, \omega \log \frac{|\omega|}{\Lambda} \,, \qquad \Sigma_\text{MFL}''(\omega) = - \frac{\lambda \pi}{2} |\omega| \,.
\ee
These obey the Kramers-Kronig relations (including a semicircle at $|z| = \Lambda$ in the contour). The parameters $\{{\mathcal A},v_F^\star,\lambda,\Lambda\}$ can vary around the Fermi surface. Indeed the MFL form may only pertain over a region of the Fermi surface. We will remain at a fixed $\bk_F$. An important property of (\ref{eq:gr2}) is that the only singular $k_\perp$ dependence is in the dispersion about the Fermi surface.

\subsection{Hertz liquid}
\label{sec:hertz}

A Hertz liquid arises from scattering of fermions by a critical boson \cite{PhysRevB.14.1165, Sachdev:2011fcc}. It is theoretically challenging to control this scattering in two-dimensional systems. Nonetheless, we can view the Hertz liquid as a phenomenological low-energy model that generalises the MFL self-energy to a continuous exponent $\a$ (cf.~\cite{Faulkner:2009wj, Reber2019, Smit2024}). That is
\be\label{eq:gr3}
G_\text{Hertz}^R(z,\bk) = \frac{{\mathcal A}}{z - v_F^\star k_\perp - \Sigma_\text{Hertz}(z)} \,,
\ee
now with
\be\label{eq:se2}
\Sigma_\text{Hertz}(z) = - i \lambda \frac{\left(-i z\right)^{\a}}{{\Lambda}^{\a-1}} + \ocal\left(\frac{z^2}{\Lambda} \right) \,.
\ee
The factors of $\Lambda$ make the coupling $\lambda$ dimensionless.
A MFL arises in the limit $\a \to 1$ (with $\Lambda$ fixed take ${\mathcal A} \approx \frac{\a-1}{\lambda_\text{MFL}} {\mathcal A}_\text{MFL}$, $v_F^\star \approx \frac{\a-1}{\lambda_\text{MFL}} v^\star_{F\,\text{MFL}}$ and $\lambda \approx - 1 + \frac{\a-1}{\lambda_\text{MFL}}$) whereas $\a = 2$ is a conventional Fermi liquid. In general, $\a$ can take any positive value. As for the MFL above, we have absorbed any linear-in-$z$ terms in the self-energy into a rescaling of the bare $z$ term, leading to the weight ${\mathcal A} \neq 1$ in (\ref{eq:gr3}). We will focus on $\a < 2$ so that the leading low-energy behaviour of the self-energy is non-Fermi liquid. This restriction also ensures that higher order analytic terms in the self-energy (\ref{eq:se2}) can be neglected for $|z| \ll \Lambda$. At real frequencies $\omega$ we now have
\be
\Sigma_\text{Hertz}'(\omega) = - \lambda \sin\left(\frac{\a \pi}{2}\right) \text{sgn}(\omega) \frac{|\omega|^\a}{\Lambda^{\a-1}} \,, \qquad \Sigma_\text{Hertz}''(\omega) = - \lambda \cos\left(\frac{\a \pi}{2}\right) \frac{|\omega|^\a}{\Lambda^{\a-1}} \,, \label{eq:H3}
\ee
which again obey the Kramers-Kronig relations. 
The requirement that $\Sigma'' < 0$ fixes the sign of $\l$. For $\a \in (0,1)$ we have $\l>0$ while for $\a \in (1,3)$ we must have $\l = - |\l| < 0$. The parameters $\{{\mathcal A}, v_F^\star, \alpha, \l\}$ can vary around the Fermi surface and the Hertz liquid may only arise at `hot spots'. The only singular $k_\perp$ dependence in (\ref{eq:gr3}) is the dispersion about the Fermi surface.

\section{Causality bounds}
\label{sec:bounds}

The pole in the retarded Green's functions (\ref{eq:gr2}) and (\ref{eq:gr3}) must remain outside of the domain of analyticity
(\ref{eq:cons}). We will probe this constraint along the pure imaginary momenta $k_\perp = i q$. Both of the self-energies (\ref{eq:se1}) and (\ref{eq:se2}) are purely imaginary when $z = i \omega_2$, with $\omega_2 > 0$. That is, we can write $\Sigma(i \omega_2) \equiv i \sigma(\omega_2)$. The pole is therefore at
\be\label{eq:sol}
v_F^\star q = \omega_2 - \sigma(\omega_2) \,.
\ee
For the two examples considered in the previous section we have
\be
\sigma_\text{MFL}(\omega_2) = \lambda \, \omega_2 \log \frac{\w_2}{\Lambda} \,, \qquad \sigma_\text{Hertz}(\omega_2) = - \l \, \frac{\omega_2^\alpha}{\L^{\a-1}} \,.
\ee
The question is whether the pole (\ref{eq:sol}) remains outside the region (\ref{eq:cons}) enforced by Lieb-Robinson causality.
This need only be imposed over the energy and momentum scales (\ref{eq:bounds}) at which the non-Fermi liquid Green's functions are applicable.

Consider first a Hertz liquid with $1 < \alpha < 3$. These is the case in which the non-analytic self-energy is subleading at low frequencies compared to the bare $\omega$ term in (\ref{eq:gr3}) and (\ref{eq:sol}). These models are mild non-Fermi liquids and have long-lived low-energy excitations. As $\omega \to 0$ it is immediate that the condition for the pole to be outside the analytic region (\ref{eq:cons}) is
\be\label{eq:vbound}
v_F^\star < v \,.
\ee
That is, the renormalised Fermi velocity must be less than the Lieb-Robinson velocity (\ref{eq:v}). This is automatically true for the bare Fermi velocity. Interactions tend to renormalise the Fermi velocity down and hence (\ref{eq:vbound}) is a reasonable expectation, which we have now proven within the context of this Hertz liquid.

\begin{figure}[h]
\centering
\includegraphics[width=0.5\textwidth]{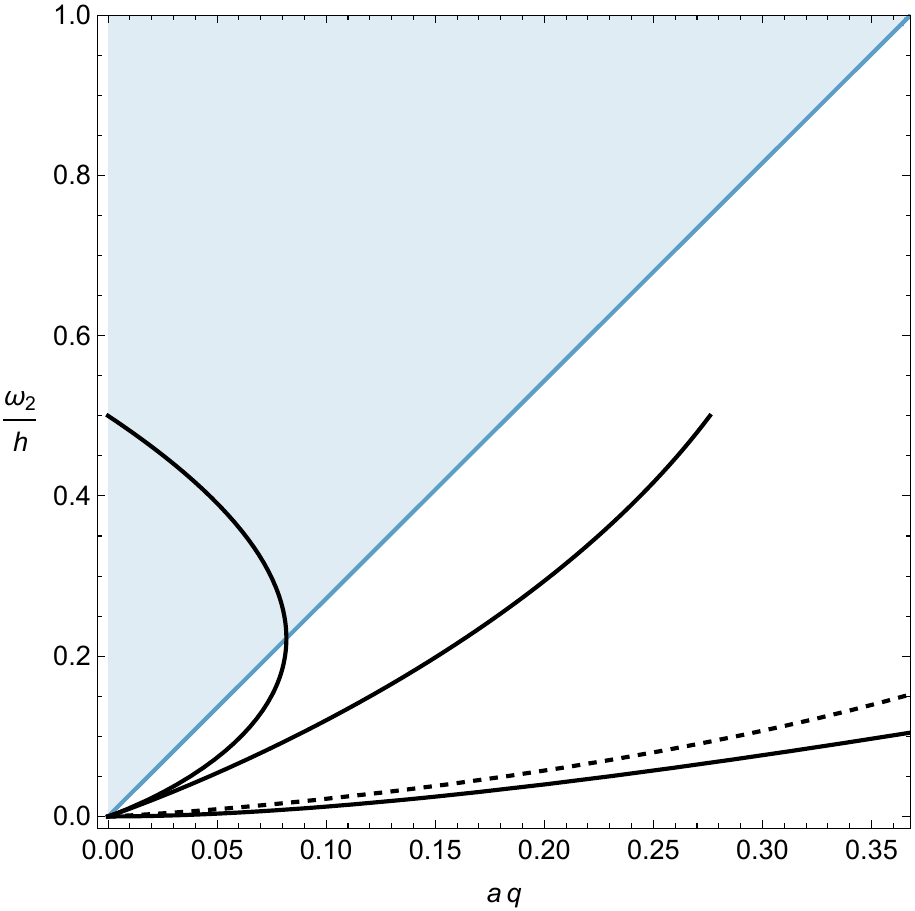}
\caption{Motion of the pole (\ref{eq:sol}) for various models over the range $0 \leq \omega_2 \leq \Lambda$. All models are shown with the illustrative values $\Lambda = \frac{1}{2} h$ and $v_F^\star = \frac{1}{3}v$. The dashed line is a MFL with $\lambda = 1$. The solid black lines are Hertz liquids. From top to bottom these have $\{\alpha = \frac{3}{2}, \lambda = -1\}$, $\{\alpha = \frac{3}{2}, \lambda = -\frac{1}{2}\}$ and $\{\alpha = \frac{1}{2}, \lambda = 1\}$. The top curve enters the shaded region, which is excluded by the causality bound (\ref{eq:cons}), and therefore does not define a consistent theory.}
\label{fig:bb}
\end{figure}

Now assume that (\ref{eq:vbound}) holds. As the frequency is increased in this model (with $1 < \alpha < 3$) the location of the pole (\ref{eq:sol}) starts to move towards the region of analyticity. This is because $\lambda = - |\lambda|$ is negative, and hence $\sigma(\omega_2)$ is positive in this case. The motion of the pole is illustrated in Fig.~\ref{fig:bb}.
To see how far the pole gets we can evaluate its location at the cutoff $\omega_2 = \Lambda$ (a more cautious reader may wish to set instead $\omega_2 = \ep \Lambda$, for some $\ep < 1$. This will weaken the bound below by a factor of $\ep^\a$). From (\ref{eq:sol}) the cutoff is reached at momentum $q_\Lambda$ given by $v_F^\star q_\Lambda = \Lambda (1 - |\lambda|)$. For the theory to be consistent with causality, the point $(q_\Lambda,\Lambda)$ must be outside of the analytic region (\ref{eq:cons}).
Typically the effective theory (\ref{eq:gr3}) will pertain at scales below the microscopic energy scale, so that $\Lambda \lesssim h$. This is a plausible but inessential additional assumption that simplifies the condition for causality in (\ref{eq:cons}) to $\Lambda < v q_\Lambda$. It follows that a consistent Hertz liquid must obey
\be\label{eq:final}
|\lambda| < 1 - \frac{v_F^\star}{v} \,. \qquad\qquad (\text{$1 < \a < 3$})
\ee
This is our second main result: a rigorous upper bound on the dimensionless coupling constant of a Hertz liquid. Fig.~\ref{fig:bb} shows two instances of Hertz liquids with $1 < \a < 3$, one that respects the bound (\ref{eq:final}) and one which does not.

Now consider the MFL and the Hertz liquid with $\a < 1$. In these models the self-energy is strongly singular at low frequencies and dominates the dispersion of the pole in (\ref{eq:sol}). As $\omega \to 0$ the pole is well outside the analytic region and there is no constraint on $v_F^\star$, as seen in Fig.~\ref{fig:bb}. Let us assume in the first instance that (\ref{eq:vbound}) nonetheless holds, i.e.~that the Fermi velocity is below the Lieb-Robinson velocity, as is typically expected. With this assumption we have
\be\label{eq:safe}
\omega_2 < \omega_2 - \sigma(\omega_2) = v_F^\star q < v q \,. \qquad\qquad (\text{MFL or $\a < 1$})
\ee
The first inequality here follows for the MFL from the fact that the log is negative for $\omega_2 < \Lambda$ and for the $\a < 1$ Hertz liquid from the fact that $\lambda$ is positive. The bound (\ref{eq:safe}) implies that the pole is always outside of the region of analyticity in these models, and there is no constraint on the coupling from causality.

As a consistency check we can verify that the coupling bound (\ref{eq:final}) is automatically obeyed in the limit in which a Hertz liquid with $1 < \alpha$ becomes a marginal Fermi liquid. The appropriate $\alpha \to 1$ limit was given in \S\ref{sec:hertz}. Indeed, (\ref{eq:final}) holds in this limit whenever $v^\star_{F\,\text{MFL}} < v$.

An intuitive, but imprecise, understanding of the results we have just obtained can be gained by defining a group velocity at real frequencies,
\be\label{eq:vg}
v_\text{g}(\omega) \equiv \frac{d\omega}{dk_\perp} = \frac{v_F^\star}{1 - \pa_\omega \Sigma'(\omega)} \,.
\ee
Requiring $v_\text{g} < v$ at all $\omega \leq \Lambda$, so that the group velocity never exceeds the Lieb-Robinson velocity, does not 
lead to the bounds above. This is because it does not account for the broadening of the fermionic excitation, which is described by the imaginary part of the self-energy. Nonetheless, (\ref{eq:vg}) captures the basic fact that when the self-energy dominates at low energies (with $\pa_\omega \Sigma'(\omega) \to - \infty$) the group velocity  goes to zero. That is to say, low-energy excitations move increasingly slowly and this is why there is no constraint from causality in these cases.

\enlargethispage{\baselineskip}

A phenomenological Hertz liquid has been used to fit ARPES data in cuprates, with the exponent $\a$ evolving as a function of doping \cite{Reber2019, Smit2024}. The data was fit to a more complicated self-energy that is a function of temperature $T$ as well as frequency which, however, reduces to (\ref{eq:gr3}) at $T=0$. For the overdoped samples it was found that $1 < \alpha < 2$ with a coupling $\bar \lambda \approx 0.5$ that is roughly independent of $\a$. It is not straightforward to convert $\bar \lambda$ to our bounded coupling $\lambda$.\footnote{The coupling $\bar \lambda$ was defined via $\bar \Sigma'' = -\bar \lambda |\omega|^{\a}/\omega_N^{\a-1}$, with $\omega_N$ a fiducial energy scale and $\bar \Sigma = \Sigma/{\mathcal A}$, with $\Sigma$ and ${\mathcal A}$ as defined in \S\ref{sec:hertz}. Comparing to (\ref{eq:H3}) we have ${\mathcal A} \bar \lambda = -|\lambda| (\frac{\omega_N}{\Lambda})^{\a-1}\cos \frac{\pi \a}{2}$. With the published analysis of the data we are not able to solve this equation to obtain our dimensionless coupling $\lambda$, as ${\mathcal A}$ is a non-universal quantity, sensitive to physics outside the low-energy model, and must be extracted from the experimental $\bar \Sigma'$.} However, the measured $\bar \lambda$ remains constant through the MFL value of $\a=1$. As explained above, a Hertz liquid 
that continuously crosses over to a MFL is automatically causal. In this sense, the data is consistent with our bound.

Finally, we can see what happens if we allow $v_F^\star > v$. 
We have noted that this is potentially fine within the MFL or $\a < 1$ effective theories because the strongly non-analytic self-energy means that the pole is not dispersing linearly at the lowest energy scales. That is to say, at the lowest energies there is no intrinsic velocity in the model and nothing is actually moving at the speed $v_F^\star$. Repeating the logic leading to (\ref{eq:final}) above, we can ask how far the pole gets at the cutoff frequency $\omega_2 = \Lambda$. Requiring that the pole remains outside the region of analyticity shows that the MFL is inconsistent with $v_F^\star > v$ while for the Hertz liquid we obtain the constraint
\be\label{eq:fast}
\frac{v_F^\star - v}{v} < \lambda \,. \qquad \qquad (\a < 1)
\ee
The bound (\ref{eq:fast}) states that $v_F^\star$ can be greater than $v$ in these cases, but not too much so.

\section{Bound on the quasiparticle residue}
\label{sec:res}

The bounds in the previous \S\ref{sec:bounds} followed from requiring the pole in the Green's function to be outside of the analytic domain.
In this section we will use (\ref{eq:grab}) to obtain an explicit upper bound on the Green's function within the domain of analyticity. 
We will see that this upper bound in turn implies an upper bound on the weight, or `quasiparticle residue', ${\mathcal A}$ of the non-Fermi liquid. We will focus on imaginary frequencies $\omega_2 < e h$ --- any low-energy effective description will be within this regime. For these frequencies the integrand in the first term in (\ref{eq:grab}) is maximised, as we described below (\ref{eq:exp}), at the upper endpoint of that integral. Thus we obtain
\begin{align}
\left| G^R(\omega,\bk) \right|  & < ||\ocal ||^2 \sum_\a \left[K \int_0^{\frac{x_\a}{eah}} dt \, e^{- \frac{\omega_2 x_\a}{e a h}} + 2 \int_{\frac{x_\a}{eah}}^\infty dt \, e^{- \omega_2 t} \right] e^{q x_\a} \\
& = ||\ocal ||^2 \sum_\a \Bigg[\frac{K}{eh} \frac{x_\a}{a}  + \frac{2}{\omega_2}  \Bigg] e^{-\left(\frac{\omega_2}{e h} - aq\right)\frac{x_\a}{a}} \,. 
\label{eq:ll}
\end{align}
For generality we have re-introduced the norm $||\ocal||$ of the local operator whose Green function is being bounded. The fermionic Green's functions we have been considering have $||\psi_0|| = 1$.

It is possible to upper bound the sum over the lattice in (\ref{eq:ll}) by a simple expression. Noting that $K \leq 2$ and $\frac{1}{\omega_2} \leq \frac{1}{\omega_2 - e a h q}$ then from (\ref{eq:ll})
\begin{align}
\left| G^R(\omega,\bk) \right|  
& < \frac{2 ||\ocal ||^2}{\omega_2 - e ah q} \sum_\a \Bigg[ \left(\frac{\omega_2}{eh}-aq\right)\frac{x_\a}{a}+1\Bigg] e^{-\left(\frac{\omega_2}{e h} - aq\right)\frac{x_\a}{a}}  \\
& \leq \frac{2 c_d (eh)^d||\ocal ||^2}{(\omega_2 - e a h q)^{d+1}} \,. \label{eq:bbb}
\end{align}
We have verified the second step numerically, using $0 < \frac{\omega_2}{eh} - a q \leq \frac{\omega_2}{eh} < 1$, as we are restricting attention to $\omega_2 < eh$. The constants $c_d = \sum_{\vec n \in \Z^d} (|\vec n| + 1)e^{-|\vec n|} \approx  \{4.005, 18.86, 100.6\}$ for $d=1,2,3$, respectively. As was to be expected, the bound becomes weaker upon approach to the low-frequency Lieb-Robinson lightcone $\omega_2 = e a h q = v q$. For simplicity we focus on $q=0$.

Applied to our non-Fermi liquid Green's functions (\ref{eq:gr2}) and (\ref{eq:gr3}), the bound (\ref{eq:bbb}) becomes a constraint on the numerator ${\mathcal A}$. Both the Green's functions and the bound concern the microscopic lattice fermion creation operator, as opposed to some renormalised operator defined within an effective field theory. These are the Green's functions measured by ARPES. With this in mind, in general there will be an `incoherent' spectral weight that should be added to the low-energy Green's functions (\ref{eq:gr2}) and (\ref{eq:gr3}). 
This additive contribution will be essentially constant at low energies and does not affect the analyticity arguments of the previous \S\ref{sec:bounds}. 
However, it contributes to the absolute value of the Green's function in the bound (\ref{eq:bbb}). Our bound on ${\mathcal A}$ in (\ref{eq:Ab}) below  assumes that the incoherent contribution is negligible compared to the non-Fermi liquid peaks in (\ref{eq:gr2}) and (\ref{eq:gr3}), at the energies where these pertain.

Using the Green's functions (\ref{eq:gr2}) and (\ref{eq:gr3}) in the bound (\ref{eq:bbb}) one finds that
the right hand side of (\ref{eq:bbb}) decreases with frequency while the left hand side increases. The strongest constraint is therefore from the upper limit $\omega_2 = \Lambda < e h$, giving our third main result that
\be
{\mathcal A} \leq 2 \, c_d \left(1 - \frac{\sigma(\Lambda)}{\Lambda} \right) \left(\frac{e h}{\Lambda} \right)^d \,.\label{eq:Ab}
\ee
Here we set $||\psi_0|| = 1$. Recall that $\sigma(\omega_2)$ was introduced in \S\ref{sec:bounds}. For the MFL, Hertz liquid with $\a < 1$ and Hertz liquid with $\a > 1$ we have, respectively, $\frac{\sigma(\Lambda)}{\Lambda} = \{0, -\lambda,|\lambda|\}$.
In the first two of these cases the bound on ${\mathcal A}$ is weaker than the general expectation that ${\mathcal A} \leq 1$. Even in these cases, however, it is notable that (\ref{eq:Ab}) is obtained purely from causality considerations. In the case of a Hertz liquid with $\a > 1$ the bound (\ref{eq:Ab}) becomes a nontrivial constraint on ${\mathcal A}$ when the pole in the Green's function at $\omega_2 = \Lambda$ approaches the boundary of the analytic region. In that case $1 - |\lambda| \approx v_F^\star/v$. Often $v_F^\star/v \ll 1$ and then the right hand side of (\ref{eq:Ab}) becomes small. Mathematically, the denominator in the $q=0$ Green's function becomes small in this limit and thus the residue is forced to be small in order to obey the bound on the full Green's function.

\section{Discussion}
\label{sec:disc}

In this paper we have shown that Lieb-Robinson causality implies a domain of analyticity (\ref{eq:cons}) and an upper bound (\ref{eq:bbb}) on the Green's function within that domain. We have used these facts to bound the coupling constant and quasiparticle residue of certain non-Fermi liquids.
These bounds are experimentally accessible and existing ARPES data is consistent with them.

The mathematical handle on lattice causality that we have developed here is a key step in building a quantum lattice bootstrap program of the kind envisioned in \cite{Chowdhury:2025dlx}, in analogy with endeavors such as the S-matrix \cite{ Paulos:2016fap, Paulos:2016but, Paulos:2017fhb} or conformal \cite{Rattazzi:2008pe, El-Showk:2012cjh, Poland:2018epd} bootstraps. An important question for future work is whether there is a bootstrap bound on the coupling constant for the case of a marginal Fermi liquid. A heuristic argument for a such a bound was given in \cite{RevModPhys.94.041002} and, as noted there, it would imply a Planckian bound for an important class of strange metallic systems.

The Lieb-Robinson causality constraints we have derived also apply to the Green's functions of conserved densities and currents, leading to bounds on transport. A lightcone bound on the diffusivity was proposed in \cite{Hartman:2017hhp}, and made sharp for relativistic causality in \cite{Heller:2022ejw, Heller:2023jtd} and for open lattice systems in \cite{Han:2018hlj}. The analyticity properties that we have established in the present paper mean that the bounds in \cite{Heller:2022ejw, Heller:2023jtd} extend to diffusion in local lattice systems (as we noted preemptively in \cite{Chowdhury:2025dlx}). We will now show that the upper bound (\ref{eq:bbb}) can be used to upper bound the conductivity of any local lattice model that is described by the Drude model at low energies. The conductivity is closely related to the diffusivity but is a distinct concept. Indeed, we will be able to set ${\bf k} = 0$ to obtain a conductivity bound and hence the only assumption for the low energy description is that the optical response is described by the Drude model \cite{armitage}.

The Green's function for the current density operator $j$ at ${\bf k} = 0$ is given in Drude theory by
\be\label{eq:drude}
G^R_{jj}(\omega) = \frac{i \omega \sigma_o}{1 - i \omega \tau} \,.
\ee
Here $\sigma_o$ is the dc conductivity and $\tau$ is the current relaxation lifetime. Setting $\omega = i \omega_2$ and using (\ref{eq:bbb}) gives an upper bound on the dc conductivity
\be\label{eq:sb}
\sigma_o \leq \frac{2 \, c_d ||j_0||^2}{a^d} \frac{1 + \omega_2 \tau}{\omega_2^2}  \left(\frac{eh}{\omega_2}\right)^d \,.
\ee
The factor of $a^d$ appears due to converting the density operator $j$ into the on-site operator $j_0$, see the recent discussion in \cite{Chowdhury:2025dlx}. The right hand side of (\ref{eq:sb}) is a decreasing function of $\omega_2$ and hence the strongest bound is obtained at large $\omega_2 = \Lambda < eh$. Here $\Lambda$ is the cutoff frequency for the validity of the Drude model.

To get a sense for the consequences of (\ref{eq:sb}), we can estimate $||j_0|| \sim a h$ and suppose that the Drude peak extends up to a finite fraction of the microscopic scale, so that $\Lambda \sim h$. Furthermore, the Drude model requires $1 \lesssim h \tau$ in order for the relaxation time to be within the regime of validity of the model. Then, up to numerical factors, we would have from (\ref{eq:sb}) that
\be\label{eq:bf}
\sigma_o \lesssim a^{2-d} h \tau \,.
\ee
The usual Drude formula for conduction by well-defined quasiparticles is $\sigma_o \sim k_F^{d-2} E_F \tau$. Here $E_F \sim k_F v_F$ is the Fermi energy. For a large Fermi surface with $k_F \sim \pi/a$ we see that (\ref{eq:bf}) becomes $v_F \lesssim v$. This is reassuringly consistent.
It should be emphasised, however, that from a low-energy perspective the two parameters $\sigma_o$ and $\tau$ in (\ref{eq:drude}) are logically independent. The causality bound (\ref{eq:bf}) can then be read as stating that, whatever the microscopic mechanism for charge transport is, the conductivity cannot be parametrically larger than the value expected from classical Drude theory. Essentially the same conclusion is reached from the causality upper bound on diffusivity \cite{Hartman:2017hhp}, again demonstrating a consistency between the various facets of Lieb-Robinson causality.

\section*{Acknowledgements}

This work has been partially supported by STFC consolidated grant ST/T000694/1. SAH is partially supported by Simons Investigator award \#620869. SDC \& AH are supported by ``Exotic High Energy Phenomenology'' (X-HEP), a project funded by the European Union -- Grant Agreement n.~101039756 (PI: J.~Elias~Mir\'o). Views and opinions expressed are however those of the author(s) only and do not necessarily reflect those of the European Union or the ERC Executive Agency (ERCEA). Neither the European Union nor the granting authority can be held responsible for them.

\appendix

\section{Lieb-Robinson bound}
\label{app:lr}

This appendix outlines a proof of the Lieb-Robinson bound, roughly following \cite{Hastings:2010vzr, Chen_2023}. This will lead to (\ref{eq:C2}) in the main text. In order to obtain explicit expressions, we consider a Hamiltonian on a $d$-dimensional square lattice, with nearest-neighbour interactions all given by the same local Hamiltonian $H$. The local terms in the Hamiltonian can be thought of as living on the edges of the lattice, connecting pairs of vertices. If we wish to keep track of the edge $\vep$ in question, we will write $H_\vep$. In addition to the operator norm $C(t,\bx_\a)$ defined in (\ref{eq:operator}) we will need to introduce the quantity
\be\label{eq:ce}
C_\vep(t) \equiv \frac{||[H_\vep(t),\psi_0(0)]||}{||H||} \,.
\ee
The time dependence here is generated using the full Hamiltonian $H_\text{full} = \sum_\vep H_\vep$, while $H_\vep$ is the local Hamiltonian on a single edge. Note that $||H_\vep|| = ||H||$.

From the definition (\ref{eq:ce}) a sequence of deceptively simple manipulations \cite{Hastings:2010vzr, Chen_2023} imply that
\be\label{eq:c1}
C(t,\bx_\a) \leq C(0,\bx_\a) + 2 ||H|| \sum_{\bx_\a \in \vep} \int_0^t C_\vep(s)\, ds \,,
\ee
where the sum is over all edges containing the vertex $\bx_\a$. Furthermore, one has that
\be\label{eq:cc}
C_\vep(t) \leq C_\vep(0) + 2 ||H|| \sum_{\vep' \cap \vep, \vep \neq \vep'} \int_0^t C_{\vep'}(s)\, ds \,.
\ee
Here the sum is over all edges $\vep'$ that share a vertex with $\vep$, but do not coincide with $\vep$. This final exclusion arises in the derivation of (\ref{eq:cc}) from the commutator $[H_\text{full},H_\vep] = \sum_{\vep' \cap \vep, \vep \neq \vep'} [H_{\vep'},H_\vep]$. The use of the two separate inequalities (\ref{eq:c1}) and (\ref{eq:cc}) accounts for the fact that $C(t,\bx_\a)$ involves an anticommutator while $C_\vep(t)$ involves a commutator. This separation also allows us to avoid taking a supremum over operators that is commonly used in derivations of the bound.

We can iterate (\ref{eq:cc}) by using the left hand side in the integral on the right hand side. This produces a sequence of edges $\vep,\vep_1,\vep_2,\cdots$. For each iteration, the first term on the right hand side of (\ref{eq:cc}) is zero unless the edge $\vep_k$ is adjacent to the origin. Let $r \equiv D(\bx_\a)$ be the Manhattan norm of $\bx_{\a}$. This is the minimal number of edges that must be traversed to get from the origin to $\bx_\a$. A given sequence of edges $\vep,\vep_1,\vep_2,\cdots$ defines path in the lattice. It follows that $C_{\vep_k}(0) = 0$ in (\ref{eq:cc}) for $k < r-1$. Thus, by iterating we obtain the bound
\begin{align}
C_\vep(t) & \leq (2 ||H||)^{r-1} \sum_{\vep_1 \cdots \vep_{r-1}} \int_0^t ds_1 \cdots \int_0^{s_{r-2}} ds_{r-1} C_{\vep_{r-1}}(s_{r-1}) \\
& \leq 2 \frac{(4(2d-1) t ||H||)^{r-1}}{(r-1)!} \,. \label{eq:bb2}
\end{align}
The second line uses the bound $C_{\vep_{r-1}}(s_{r-1}) \leq 2$ and the fact that, given an edge $\vep$ on a square lattice, there are $2(2d-1)$ edges $\vep'$ obeying the conditions in the sum (\ref{eq:cc}).

Using (\ref{eq:bb2}) in (\ref{eq:c1}) gives
\be\label{eq:Cfin}
C(t,\bx_\a) \leq K \frac{(ht)^r}{r!} \,.
\ee
Here we introduced
\be
h \equiv 4(2d-1) ||H|| \,, \qquad K \equiv \frac{2d}{2d-1}  \,,
\ee
and used the fact that there are $2d$ edges connecting to the vertex $\bx_\a$. The first term in (\ref{eq:c1}) is zero unless $\bx_\a$ coincides with the origin. Thus (\ref{eq:Cfin}) holds for all $r \geq 1$.
Recalling that we always have $C(t,\bx_\a) \leq 2$ we may strengthen the bound (\ref{eq:Cfin}) to
\be\label{eq:min}
C(t,\bx_\a) \leq \min\left(2, K \frac{(h t)^{D(\bx_\a)}}{D(\bx_\a)!}  \right) < 
\min\left(2, K \left(\frac{e h t}{D(\bx_\a)}\right)^{D(\bx_\a)} \right) \,.
\ee
Here we re-introduced the Manhattan norm explicitly, setting $r \equiv D(\bx_\a)$.
The second inequality in (\ref{eq:min}) uses a standard bound on factorial.

For the arguments in the main text to go through we will want to use the final expression in (\ref{eq:min}) as the upper bound for $x/a > eht$, and the upper bound of $2$ for $x/a < eht$. This restriction allows us to get rid of the Manhattan norm in (\ref{eq:min}). The Manhattan norm is necessarily greater than the Euclidean norm, $D(\bx) \geq x/a$ with $x = |\bx|$, and furthermore the function $f(x) = (e x_o/x)^x$ is monotonically decreasing for $x > x_o$. That is, we can replace $D(\bx)$ by $x/a$ in the final expression in (\ref{eq:min}) whenever $x/a > h t$. This is a weaker condition than our requirement that $x/a > eht$. That is, we can take
\be
C(t,\bx) <
\left\{
\begin{array}{cl}
K \left(\frac{e h t}{x/a}\right)^{x/a} & \text{for $x/a > eht$}  \\
2 & \text{for $x/a < e h t$}
\end{array}
\right. \,,
\ee
which is (\ref{eq:C2}) in the main text.

\providecommand{\href}[2]{#2}\begingroup\raggedright\endgroup

\end{document}